\documentclass[12pt]{article}

\usepackage[koi8-r]{inputenc}
\usepackage[english]{babel}

\usepackage{amsmath}
\usepackage{epsf}
\usepackage{epsfig}
\usepackage{here}
\usepackage{amssymb}
\usepackage{graphicx}
\def\sw{\text{sw}}
\def\m{\text{m}}
\def\s{\text{s}}

\makeatletter

\def\lsim{\compoundrel<\over\sim}
\def\compoundrel#1\over#2{\mathpalette\compoundreL{{#1}\over{#2}}}
\def\compoundreL#1#2{\compoundREL#1#2}
\def\compoundREL#1#2\over#3{\mathrel
         {\vcenter{\hbox{$\m@th\buildrel{#1#2}\over{#1#3}$}}}}
\makeatother

\begin{document}

\begin{center}

    {\Large\bf Gravitational waves  from phase transition 
    in split NMSSM}
    \\
    \vspace{0.8cm}
    \vspace{0.3cm}
    S.~V.~Demidov$^{a,b,}$\footnote{{\bf e-mail}: demidov@ms2.inr.ac.ru}, 
    D.~S.~Gorbunov$^{a,b,}$\footnote{{\bf e-mail}: gorby@ms2.inr.ac.ru},
    D.~V.~Kirpichnikov$^{a,}$\footnote{{\bf e-mail}: kirpich@ms2.inr.ac.ru}
    \\
    
    $^a${\small{\em 
        Institute for Nuclear Research of the Russian Academy of Sciences, }}\\
      {\small{\em
          60th October Anniversary prospect 7a, Moscow 117312, Russia
      }
      }
      \\
$^{b}${\small{\em
Moscow Institute of Physics and Technology,
}}\\
{\small{\em
Institutsky per. 9, 
  Dolgoprudny 141700, Russia
}}\\
  \end{center}
  \begin{abstract}
We discuss gravitational wave signal from the strongly first order electroweak
phase transition in the split NMSSM. We find that for sets of
parameters predicting successful electroweak baryogenesis the gravitational wave
signal can be within the reach of future experiments LISA, BBO and
Ultimate DECIGO.
  \end{abstract}

The phenomenological searches for physics beyond the Standard Model
(SM)  with gravitational wave (GW) interferometers can significantly
improve our understanding of matter properties and the early Universe 
evolution. Recent LIGO and VIRGO detection of the GW signal 
from binary black hole mergers
\cite{Abbott:2016blz,Abbott:2016nmj,Abbott:2017vtc}
opens up opportunities in the field for the future space-based
GW interferometers LISA\,\cite{Seoane:2013qna,Audley:2017drz}, DECIGO\,\cite{Seto:2001qf,Kudoh:2005as} and
BBO\,\cite{Harry:2006fi}. Among numerous GW signals
originated from different sources (for a recent review see,
e.g.,\,\cite{Cai:2017cbj}) one can single out gravitational 
waves from the first order electroweak phase transition (EWPT). It is
known that this transition is a crossover in the SM\,\cite{Kajantie:1996mn,Csikor:1998eu}. However,
new scalar degrees of freedom introduced in various extensions of the SM 
at the electroweak scale  
can turn this transition to
the first order phase transition. This possibility is quite
interesting because on the one hand it allows for
generation of the observed baryon asymmetry of the Universe (BAU). On
the other hand, it requires a change of physics around TeV scale which
could be potentially probed by the ongoing experiments at the
Large Hadron Collider (LHC). GW signal from the EWPT would be an independent
signature of such new physics models. In particular, it is important
for testing the blind spots of the LHC searches. The GW signal from the EWPT was
recently studied within the Next-to-Minimal Supersymmetric Standard
Model (NMSSM)~\cite{Bian:2017wfv}, in the model with
singlet extension of the Higgs
sector~\cite{Kakizaki:2015wua,Beniwal:2017eik}, in 6D operator extension  
of the SM~\cite{Huang:2016odd}, and  in the scenarios with a hidden QCD-like
sector of the SM~\cite{Aoki:2017aws, Anand:2017kar}.  

In this paper we  estimate the GW signal from EWPT within the
framework of split  NMSSM~\cite{Demidov:2006zz,Demidov:2016wcv}. This
model provides with the strongly first order EWPT due to specific
singlet extension of the Higgs sector. Percolation of the new phase
bubbles eventually produces stochastic background of GWs, which is not
flat, but exhibits a characteristic peak. We find that for the benchmark
points in parameter space of this model with successful electroweak
baryogenesis the GW signals are expected to be right  
in the sensitivity region of
the proposed space-based interferometers LISA, BBO, DECIGO. Their
observation would provide with independent evidence for the new
physics playing important role in the early Universe.

We start with recalling basic properties of the split
NMSSM~\cite{Demidov:2006zz,Demidov:2016wcv}. This model is an
extension of split MSSM~\cite{ArkaniHamed:2004fb,Giudice:2004tc} with
an additional singlet superfield. The latter contains CP-even $S$ and
CP-odd $P$ scalars as well as singlet fermion $\tilde{n}$. Particle
spectrum in this model is splitted: all the scalars except for the
lightest Higgs boson $h$ and the singlet scalars are heavy with their
masses around a splitting scale $M_S$. To be consistent with the
measured Higgs boson properties this scale should be about
10--20~TeV~\cite{Demidov:2016wcv}. Other superpartners (neutralinos,
charginos, etc) are supposed to
be around (sub)TeV scale. 
The low energy Lagrangian can be obtained by integrating out the heavy
particles and is presented in~\cite{Demidov:2006zz,Demidov:2016wcv}.

It was shown in Ref.\,\cite{Demidov:2016wcv} that in the framework of split
NMSSM  the observed baryon asymmetry of the Universe (BAU) can be 
generated during the strongly first order EWPT. The strengthening of the
PT with respect to the SM is a result of modification of the scalar
potential, which is more complicated in the model due to presence
  of new scalar fields.  
The source of CP-violation in this model  is associated with complex 
chargino mass matrix and in particular with the effective
$\mu$-parameter~\cite{Demidov:2006zz}. As the split NMSSM provides
strongly first order PT it can be potentially probed
with GW echo from colliding bubbles of the new phase. 
In what follows we consider the benchmark points in the parameter
space of the model marked as {\it Setup 1} and {\it Setup 2} in
Ref.~\cite{Demidov:2016wcv}.
Apart from predicting the
observed baryon asymmetry of the Universe,  they satisfy current
experimental constraints from LHC data and the EDM bounds and suggest
the lightest neutralino as a viable DM candidate.  

The dynamics of cosmological first-order PT via tunneling of the
scalar fields from false to true vacuum  has been thoroughly studied 
in literature, for review see e.g.
Refs.~\cite{Espinosa:2010hh,Weir:2017wfa,Jinno:2017ixd}. The 
corresponding  probability of bubble nucleation per unit Hubble
space-time volume at temperature $T$ is given by  
\begin{equation}
  \label{probability}
P \simeq M_{Pl}^4 / T^4  \cdot \exp(-S_3(T)/T), 
\end{equation}
where $S_3$ is the three dimensional Euclidean action or,
equivalently, free energy of the scalar bubble configuration
\begin{equation}
  \label{action}
S_3(T) = 4\pi \int \limits_0^{\infty} dr \, r^2 \left[ 
\frac{1}{2} \left( \frac{dh }{dr}\right)^2+
\frac{1}{2} \left( \frac{dS }{dr}\right)^2+
\frac{1}{2} \left( \frac{dP }{dr}\right)^2+V_T^{\text{eff}}(h,S,P)
 \right].
\end{equation} 
with $V_T^{\text{eff}}(h,S,P)$ being a one-loop effective  potential  at 
the finite temperature~\cite{Demidov:2006zz}. The bubbles of new phase 
nucleate when $P\sim 1$, and eq.\,\eqref{probability} yields  
$S_3/T \sim 4 \log (M_{Pl}^*/T) \sim 150$ 
at the typical electroweak temperature $T \simeq 100$\,GeV. The accurate 
calculation\,\cite{Anderson:1991zb} reveals the following nucleation 
condition for the free energy and nucleation temperature $T_c$:  
\begin{equation}
130<S_3(T_c)/T_c<140\,.
\label{NuclCond}
\end{equation}
Note that here we are interested in one-step PT for split
NMSSM. However, there is another option:  the first order PT can go in two steps
in  models with singlet extension of the Higgs
sector if there exists a metastable vacuum of the effective
potential. Moreover, such multi-step PT is also associated with a
GW signal which can be measured by LISA and DECIGO, see
e.g. Refs.\,\cite{Huber:2015znp,Kang:2017mkl}.  We do not
  consider this exotic option in the split NMSSM, which requires
  special investigation of the electroweak baryogenesis.

We use semianalytical results for the spectrum of GW from the first
order EWPT which were obtained from numerical simulations.  The
GW are originated mainly from three sources (see
e.g.~\cite{Caprini:2015zlo} for a recent review): bubble 
collisions\footnote{
See also Ref.~\cite{Jinno:2016vai} for an analytical calculation of
this spectrum consistent with the numerical
results.}\,\cite{Huber:2008hg},
sound waves produced by
percolation\,\cite{Hindmarsh:2015qta}, 
and turbulent motion of plasma originated from the same
percolation\,\cite{Caprini:2009yp}. Let us introduce key
parameters which characterize the GW spectrum; hereafter we adopt the
notations of Ref.\,\cite{Caprini:2015zlo}. The rate of bubble
nucleation, i.e. the inverse
duration of the PT,  is
\begin{equation}
\beta_c \equiv  \frac{\beta}{H_c} \equiv T\frac{d}{dT}\left(\frac{S_3}{T}\right)\Bigg|_{T=T_c}
\end{equation}
where  $H_c$ is the  Hubble 
parameter at the nucleation temperature $T_c$.
The GW spectrum also depends on the ratio of latent heat $\epsilon$
released during the PT and radiation energy  density
$\rho_{\gamma}$, 
\begin{equation}
\alpha \equiv \epsilon/\rho_{\gamma}\equiv \frac{1}{\rho_\gamma}\left(\Delta  V_T^{\text{eff}} 
-\frac{T}{4}    \left(\frac{d \Delta V_T^{\text{eff}}}{d T}\right)\right)\Bigg|_{T_c}\,.
\end{equation}
Here variation 
$\Delta  V_T^{\text{eff}} =  V_T^{\text{eff}}(\phi_{symm}) -
V_T^{\text{eff}}(\phi_{broken})$ is the potential energy difference
between symmetric and broken phase.

In our case the surface of expanding bubble efficiently interacts with
particles in the plasma. This friction limits the bubble wall
velocity, which we expect to be subsonic, $v_w<c_s\equiv1/\sqrt{3}\approx
0.57$. It makes the contribution to the GW spectrum from the
  bubble collisions
negligible. At the same time, both the sound waves and the turbulence 
survive in the plasma for quite a long time after the bubble
collisions at the PT, that makes their contributions dominant.  

The GW signal is usually presented as an integrated  contribution  
 of the logarithmic wave interval to the total energy density of the present
Universe. The latter is characterized as $\Omega_{\text{GW}} h^2$,
where $\Omega_{\text{GW}}$ is a relative contribution of the GW, while
$h=0.68$ refers to the present value of the Hubble
parameter. Then the main sources of the GW signal ---
the sound waves $\Omega_{\text{sw}} h^2(f)$ and the
magnetohydrodynamic turbulence in plasma $\Omega_{\text{m}} h^2(f)$ 
--- read  
\begin{equation}
\Omega_{\text{GW}} h^2(f) =  \Omega_{\text{sw}} h^2(f)+\Omega_{\text{m}} h^2(f),
\end{equation}  
where $\Omega_{\text{sw}} h^2(f)$ and $\Omega_{\text{m}} h^2(f)$
are, respectively, given by~\cite{Caprini:2015zlo} 
\begin{equation}
\Omega_{\text{sw}} h^2(f) = 1.23\times 10^{-5} g_*^{-1/3} \beta_c^{-1}
\left( \frac{k_{\text{sw}}  \alpha}{1+\alpha}\right)^2 v_w\, S_{\text{sw}}(f), \label{OmegaSW}
\end{equation}
\begin{equation}
\Omega_{\text{m}} h^2(f) = 1.55\times 10^{-3} g_*^{-1/3} \beta_c^{-1}
\left( \frac{k_{\text{m}}  \alpha}{1+\alpha}\right)^{3/2} v_w\, S_{\text{m}}(f), \label{OmegaM}
\end{equation}
where $v_w$ is the average bubble wall velocity and $g_*$ is the
effective number of degrees of freedom in the plasma. 
Corresponding functions of frequency $f$ in Eqs.~(\ref{OmegaSW}) 
and~(\ref{OmegaM})  read as   
\begin{equation}
\label{spectra}
  S_{\text{sw}}=\left(\frac{f}{f_{\text{sw}}}\right)^3
\left(\frac{7}{4+3 (f/f_{\text{sw}})^2}\right)^{7/2}, \qquad 
S_{\text{m}}=\frac{(f/f_{\text{m}})^3}{(1+f/f_{\text{m}})^{11/3} (1+8 \pi f/h_*) }
\end{equation}       
which peak, respectively, at frequencies  
\begin{equation}
f_{\text{sw}}\simeq 1.15\,\frac{\beta_c h_*}{v_w}\,, \hskip2cm f_{\text{m}} \simeq 1.65\, \frac{\beta_c h_*}{v_w}\,,  
\end{equation}
where 
\begin{equation}
h_* \equiv 1.65\times 10^{-5}\,\mbox{Hz}\left(\frac{T_c}{100\,\mbox{GeV}}\right)\left(\frac{g_*}{100}\right)^{1/6}.  
\end{equation}
The coefficients $k_{\text{sw}}$ and $k_{\text{m}}$ entering
eqs.\,\eqref{OmegaSW} and \eqref{OmegaM} are 
fractions of the 
released vacuum energy transformed to the bulk motion
of the medium and magnetic turbulence of the plasma, respectively,
which for the case of subsonic bubble wall read (see
Refs.~\cite{Espinosa:2010hh,Caprini:2015zlo})  
\begin{equation}
k_{\text{sw}} = 
\frac{c_\s^{11/5} k_a k_b}{ \left(c_\s^{11/5} - v_w^{11/5} 
 \right)k_b +v_w c_\s^{6/5} k_a}\,, \qquad k_\m = 0.05\, k_\sw\,, 
\end{equation}
here the parameters  $k_a$ and $k_b$ are given by 
\begin{equation}
k_a= \frac{6.9 v^{6/5}_w \alpha}{1.36-0.037 \sqrt{\alpha}+\alpha}\,,
\quad k_b= \frac{ \alpha^{2/5}}{0.017 +(0.9997+\alpha)^{2/5}}\,.
\end{equation} 

\begin{figure}[!htb]
\begin{center}
\includegraphics[width=0.54\textwidth]{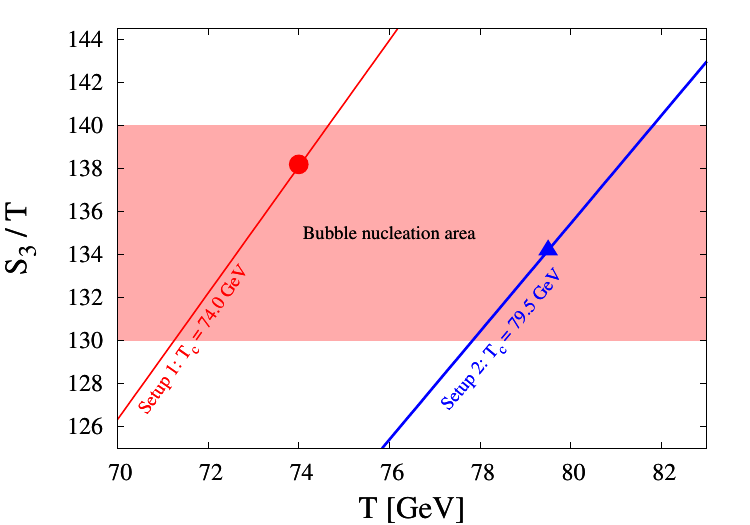}
\caption {Plot of  $S_3/T$ as a function of $T$ for the two benchmark
  points of Table\,\ref{TableOfSetup}.  
\label{figure2}}
\end{center}
\end{figure}

\begin{table}[!htb]
\begin{center}
\begin{tabular}{|c|c|c|c|c|}
\hline
$ $ & $T_c$ [GeV] & $S_3/T_c$ & $ \alpha $ & $\beta_c$ \\
\hline
{\it Setup 1} & $ 74.0$ & $138.2$ & $6.2\times 10^{-2}$ & $214$ \\
\hline
{\it Setup 2} & $79.5$ & $ 134.1$ & $4.5\times 10^{-2}$ & $200$ \\
\hline
\end{tabular}
\end{center}
\caption{Benchmark points of the EWPT  predicted in 
 the split NMSSM\,\cite{Demidov:2016wcv}.
\label{TableOfSetup}}
\end{table}
Numerical values of the parameters characterizing the EWPT for the two chosen 
benchmark models are shown in Table\,\ref{TableOfSetup}.
Namely we
present here the critical temperature $T_c$ and $S_3/T_c$ which were 
calculated previously in~\cite{Demidov:2016wcv}. Using the Euclidean
action~\eqref{action} and numerical procedure to find the bounce
solution which were employed in~\cite{Demidov:2016wcv} we numerically
calculate the derivative
$\frac{d}{dT}\left(\frac{S_3}{T}\right)\Big|_{T=T_c}$,  
as well as the bubble nucleation rate $\beta_c$, and the
parameter $\alpha$. In Fig.\,\ref{figure2} 
we show 
$S_3/T$ as a function of $T$ in the vicinity of the bubble
  nucleation temperature. The points in the Figure indicate chosen 
  moments of the phase transition for {\it Setup 1} and {\it Setup
    2}.   Let us note that the uncertainty in the nucleation
  condition~\eqref{NuclCond} results in an uncertainty of few GeV in 
  determination of the nucleation temperature. Finally, for
  parameters in Table~\ref{TableOfSetup} we  
  obtain the spectra of the GW for two values $v_w=0.1$ and $v_w=0.5$.
Precise calculation of the
  bubble wall velocity entering Eqs.~(\ref{OmegaSW}) 
and~(\ref{OmegaM}) is rather involved (see,
e.g.~\cite{Huber:2011aa,Kozaczuk:2015owa}). 
Here we consider an optimistic subsonic
case for the bubble wall velocity $v_w < 1/\sqrt{3} \approx 0.57$.
Let us note that the baryon asymmetry of the Universe depends rather
weakly~\cite{Demidov:2016wcv} on $v_w$.

\begin{figure}[!htb]
\begin{center}
\includegraphics[width=0.8\textwidth]{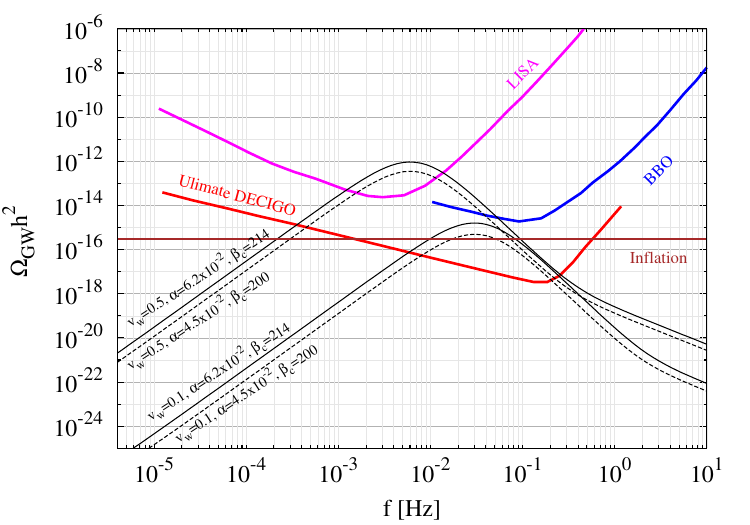}
\caption{Sensitivity of proposed GW interferometers
  \cite{Kudoh:2005as,Thrane:2013oya} and predictions of GW spectrum in
  split NMSSM for the two setups of Table\,\ref{TableOfSetup}.  Setup
  1: black solid line, $T_c=74.0$\,GeV; Setup 2: black dashed line,
  $T_c=79.5$\,GeV.  For these parameters the substantial GW signal can
  be potentially detected by LISA for the bubble wall velocity
  $v_w\simeq 0.5$.
\label{figure1}}
\end{center}
\end{figure}

  Our results for the GW spectra as well as numerical values of
  $\alpha$ and $\beta_c$ are presented in
  Fig.\,\ref{figure1}. 
The sensitivities of LISA, BBO and DECIGO space-based interferometers
to these benchmark points are also shown in Fig.~\ref{figure1}. 
The central part of the GW spectra containing a maximum is
  dominated by the acoustic wave contribution. 
  At very low frequencies both spectra~\eqref{spectra} grow as $f^3$
  but contribution from MHD turbulence dominates. At high frequencies
  the latter again takes the lead and reveal the expected Kolmogorov
  turbulent behaviour with (-5/3) power. 
One can see from Eqs.~(\ref{OmegaSW}) 
and~(\ref{OmegaM}) that the GW signal scales\footnote{
Let us note that this scaling has recently been questioned in
Ref.~\cite{Jinno:2017fby}, where an analytical approach is used
to calculate the GW spectra from sound waves.}
as $\sim \beta_c^{-1}\alpha^2$.   
This means that small values of $\beta_c$ and large $\alpha$ provide 
a higher GW signal. Therefore the GW signal in Fig.~\ref{figure1}
for Setup 1 is enhanced for slightly larger value
$\alpha=6.2\times 10^{-2}$  with respect to $\alpha=4.5\times 10^{-2}$
from Setup 2.  
Comparing the predictions with the expected sensitivities of
  different experiments we find that the most sensitive LISA
configuration~\cite{Caprini:2015zlo} can test all these setups for   
$v_w \simeq 0.5$.

Let us note that current bound on
  tensor-to-scalar perturbation ration $r\lsim 0.1$ implies that
  contribution of gravitational waves from inflation to
  $\Omega_{GW}h^2$ should be smaller than about $3\times 10^{-16}$
  assuming flat perturbation spectrum. Thus the GW signal from
  EWPT predicted within split NMSSM can be distinguished from this
  irreducible background.

Finally, let us make a brief comment on neutralino-chargino sector of
the model in context of latest results of searches for superpartners
at the LHC experiments. For the two benchmark models in question the
soft parameter $M_2$ is taken to be 1~TeV for concreteness although our
results concerning the phase transition and BAU~\cite{Demidov:2016wcv} are almost
independent of this parameter if it is considerably larger than the
critical temperature. This leaves only a tiny wino contribution in the
lighter chargino and neutralino species which decreases sensitivity of
corresponding LHC searches (see e.g.~\cite{atlas_conf}) with the
massive gauge bosons in the final state due to lower coupling of
higgsino-like fermions to the gauge bosons. We take it as an
additional motivation for the present study.

\vskip 0.3cm 
The work was supported by the RSF grant 14-22-00161.


\begin{thebibliography}{99}

\bibitem{Abbott:2016blz}
  B.~P.~Abbott {\it et al.} [LIGO Scientific and Virgo Collaborations],
  ``Observation of Gravitational Waves from a Binary Black Hole Merger,''
  Phys.\ Rev.\ Lett.\  {\bf 116} (2016) no.6,  061102.

\bibitem{Abbott:2016nmj}
  B.~P.~Abbott {\it et al.} [LIGO Scientific and Virgo Collaborations],
  Phys.\ Rev.\ Lett.\  {\bf 116} (2016) no.24,  241103
  doi:10.1103/PhysRevLett.116.241103
  [arXiv:1606.04855 [gr-qc]].
  
\bibitem{Abbott:2017vtc}
  B.~P.~Abbott {\it et al.} [LIGO Scientific and VIRGO Collaborations],
  ``GW170104: Observation of a 50-Solar-Mass Binary Black Hole Coalescence at Redshift 0.2''
  Phys.\ Rev.\ Lett.\  {\bf 118} (2017) no.22,  221101.
 
\bibitem{Seoane:2013qna} 
  P.~A.~Seoane {\it et al.} [eLISA Collaboration],
  ``The Gravitational Universe,''
  arXiv:1305.5720 [astro-ph.CO].

\bibitem{Audley:2017drz}
  H.~Audley {\it et al.},
  arXiv:1702.00786 [astro-ph.IM].

  
\bibitem{Hindmarsh:2015qta}
  M.~Hindmarsh, S.~J.~Huber, K.~Rummukainen and D.~J.~Weir,
  ``Numerical simulations of acoustically generated gravitational waves at a first order phase transition,''
  Phys.\ Rev.\ D {\bf 92} (2015) no.12,  123009.
  
\bibitem{Caprini:2009yp}
  C.~Caprini, R.~Durrer and G.~Servant,
  JCAP {\bf 0912} (2009) 024
  doi:10.1088/1475-7516/2009/12/024
  [arXiv:0909.0622 [astro-ph.CO]].
  
\bibitem{Seto:2001qf}
  N.~Seto, S.~Kawamura and T.~Nakamura,
  ``Possibility of direct measurement of the acceleration of the universe using 0.1-Hz band laser interferometer gravitational wave antenna in space,''
  Phys.\ Rev.\ Lett.\  {\bf 87} (2001) 221103.
  
\bibitem{Kudoh:2005as}
  H.~, A.~Taruya, T.~Hiramatsu and Y.~Himemoto,
  ``Detecting a gravitational-wave background with next-generation space interferometers,''
  Phys.\ Rev.\ D {\bf 73} (2006) 064006.
  

\bibitem{Harry:2006fi}
  G.~M.~Harry, P.~Fritschel, D.~A.~Shaddock, W.~Folkner and E.~S.~Phinney,
  Class.\ Quant.\ Grav.\  {\bf 23} (2006) 4887
   Erratum: [Class.\ Quant.\ Grav.\  {\bf 23} (2006) 7361].
  doi:10.1088/0264-9381/23/24/C01, 10.1088/0264-9381/23/15/008

\bibitem{Cai:2017cbj}
  R.~G.~Cai, Z.~Cao, Z.~K.~Guo, S.~J.~Wang and T.~Yang,
  doi:10.1093/nsr/nwx029
  arXiv:1703.00187 [gr-qc].
  
   

\bibitem{Kajantie:1996mn}
  K.~Kajantie, M.~Laine, K.~Rummukainen and M.~E.~Shaposhnikov,
  Phys.\ Rev.\ Lett.\  {\bf 77} (1996) 2887
  doi:10.1103/PhysRevLett.77.2887
  [hep-ph/9605288].


\bibitem{Csikor:1998eu}
  F.~Csikor, Z.~Fodor and J.~Heitger,
  Phys.\ Rev.\ Lett.\  {\bf 82} (1999) 21
  doi:10.1103/PhysRevLett.82.21
  [hep-ph/9809291].

\bibitem{Bian:2017wfv}
  L.~Bian, H.~K.~Guo and J.~Shu,
  ``Gravitational Waves, baryon asymmetry of the universe and electric dipole moment in the CP-violating NMSSM''.
 
  
\bibitem{Huang:2016odd}
  F.~P.~Huang, Y.~Wan, D.~G.~Wang, Y.~F.~Cai and X.~Zhang,
  ``Hearing the echoes of electroweak baryogenesis with gravitational wave detectors,''
  Phys.\ Rev.\ D {\bf 94} (2016) no.4,  041702.
  
\bibitem{Kakizaki:2015wua}
  M.~Kakizaki, S.~Kanemura and T.~Matsui,
  ``Gravitational waves as a probe of extended scalar sectors with the first order electroweak phase transition,''
  Phys.\ Rev.\ D {\bf 92} (2015) no.11,  115007.
  
\bibitem{Beniwal:2017eik}
  A.~Beniwal, M.~Lewicki, J.~D.~Wells, M.~White and A.~G.~Williams,
  ``Gravitational wave, collider and dark matter signals from a scalar singlet electroweak baryogenesis''.
  JHEP {\bf 1708} (2017) 108

\bibitem{Aoki:2017aws}
  M.~Aoki, H.~Goto and J.~Kubo,
  ``Gravitational Waves from Hidden QCD Phase Transition''.
  
\bibitem{Anand:2017kar}
  S.~Anand, U.~K.~Dey and S.~Mohanty,
  ``Effects of QCD Equation of State on the Stochastic Gravitational Wave Background,''
  JCAP {\bf 1703} (2017) no.03,  018.
 

\bibitem{Demidov:2006zz}
  S.~V.~Demidov and D.~S.~Gorbunov,
  ``Non-minimal Split Supersymmetry,''
  JHEP {\bf 0702} (2007) 055.

\bibitem{Demidov:2016wcv}
 S.~V.~Demidov, D.~S.~Gorbunov and D.~V.~Kirpichnikov,
 ``Split NMSSM with electroweak baryogenesis,''
 JHEP {\bf 1611} (2016) 148
 Erratum: [JHEP {\bf 1708} (2017) 080].

\bibitem{ArkaniHamed:2004fb}
  N.~Arkani-Hamed and S.~Dimopoulos,
  JHEP {\bf 0506} (2005) 073
  doi:10.1088/1126-6708/2005/06/073
  [hep-th/0405159].
\bibitem{Giudice:2004tc}
  G.~F.~Giudice and A.~Romanino,
  Nucl.\ Phys.\ B {\bf 699} (2004) 65
   Erratum: [Nucl.\ Phys.\ B {\bf 706} (2005) 487]
  doi:10.1016/j.nuclphysb.2004.11.048, 10.1016/j.nuclphysb.2004.08.001
  [hep-ph/0406088].


 
\bibitem{Espinosa:2010hh}
  J.~R.~Espinosa, T.~Konstandin, J.~M.~No and G.~Servant,
  ``Energy Budget of Cosmological First-order Phase Transitions,''
  JCAP {\bf 1006} (2010) 028.
  
\bibitem{Weir:2017wfa}
  D.~J.~Weir,
  ``Gravitational waves from a first order electroweak phase transition: a review,''
  arXiv:1705.01783 [hep-ph].
  
  
\bibitem{Jinno:2017ixd}
  R.~Jinno, S.~Lee, H.~Seong and M.~Takimoto,
  arXiv:1708.01253 [hep-ph].
  
  
\bibitem{Anderson:1991zb}
  G.~W.~Anderson and L.~J.~Hall,
  ``The Electroweak phase transition and baryogenesis,''
  Phys.\ Rev.\ D {\bf 45} (1992) 2685.

\bibitem{Huber:2015znp}
  S.~J.~Huber, T.~Konstandin, G.~Nardini and I.~Rues,
  JCAP {\bf 1603} (2016) no.03,  036
  doi:10.1088/1475-7516/2016/03/036
  [arXiv:1512.06357 [hep-ph]].

  
\bibitem{Kang:2017mkl}
  Z.~Kang, P.~Ko and T.~Matsui,
  ``Strong First Order EWPT and Strong Gravitational Waves in $Z_3$-symmetric Singlet Scalar Extension,''
  arXiv:1706.09721 [hep-ph].

\bibitem{Caprini:2015zlo}
  C.~Caprini {\it et al.},
  ``Science with the space-based interferometer eLISA. II: Gravitational waves from cosmological phase transitions,''
  JCAP {\bf 1604} (2016) no.04,  001.

\bibitem{Jinno:2016vai}
  R.~Jinno and M.~Takimoto,
  Phys.\ Rev.\ D {\bf 95} (2017) no.2,  024009
  doi:10.1103/PhysRevD.95.024009
  [arXiv:1605.01403 [astro-ph.CO]].

  
\bibitem{Huber:2008hg}
  S.~J.~Huber and T.~Konstandin,
  JCAP {\bf 0809} (2008) 022
  doi:10.1088/1475-7516/2008/09/022
  [arXiv:0806.1828 [hep-ph]].
  
  
\bibitem{Huber:2011aa}
  S.~J.~Huber and M.~Sopena,
  ``The bubble wall velocity in the minimal supersymmetric light stop scenario,''
  Phys.\ Rev.\ D {\bf 85} (2012) 103507.
\bibitem{Kozaczuk:2015owa}
  J.~Kozaczuk,
  JHEP {\bf 1510} (2015) 135
  doi:10.1007/JHEP10(2015)135
  [arXiv:1506.04741 [hep-ph]].

\bibitem{Bodeker:2009qy}
  D.~Bodeker and G.~D.~Moore,
  JCAP {\bf 0905} (2009) 009
  doi:10.1088/1475-7516/2009/05/009
  [arXiv:0903.4099 [hep-ph]].

  
\bibitem{Thrane:2013oya}
  E.~Thrane and J.~D.~Romano,
  Phys.\ Rev.\ D {\bf 88} (2013) no.12,  124032
  doi:10.1103/PhysRevD.88.124032
  [arXiv:1310.5300 [astro-ph.IM]].


\bibitem{Jinno:2017fby}
  R.~Jinno and M.~Takimoto,
  arXiv:1707.03111 [hep-ph].

  
\bibitem{atlas_conf}
  {\it ATLAS Collaboration}, 
  ``Search for electroweak production of supersymmetric particles in
  the two and three lepton final state at
  $\boldmath{\sqrt{s}=13\,}$TeV with the ATLAS detector'', ATLAS-CONF-2017-039
  

\end{thebibliography}
\end{document}